# Sparse Direction of Arrival Estimation Method Based on Vector Signal Reconstruction with a Single Vector Sensor

Jiabin Guo[1]

[1] *Acoustic Science and Technology Laboratory, Harbin Engineering University, Harbin, 150001, China*

This study investigates the application of single vector hydrophones in underwater acoustic signal processing for Direction of Arrival (DOA) estimation. Addressing the limitations of traditional DOA estimation methods in multi-source environments and under noise interference, this research proposes a Vector Signal Reconstruction (VSR) technique. This technique transforms the covariance matrix of single vector hydrophone signals into a Toeplitz structure suitable for gridless sparse methods through complex calculations and vector signal reconstruction. Furthermore, two sparse DOA estimation algorithms based on vector signal reconstruction are introduced. Theoretical analysis and simulation experiments demonstrate that the proposed algorithms significantly improve the accuracy and resolution of DOA estimation in multi-source signals and low Signal-to-Noise Ratio (SNR) environments compared to traditional algorithms. The contribution of this study lies in providing an effective new method for DOA estimation with single vector hydrophones in complex environments, introducing new research directions and solutions in the field of vector hydrophone signal processing.

## I. INTRODUCTION AND FIRST-LEVEL HEADINGS

In array signal processing, the estimation of the Direction of Arrival (DOA) is a crucial component of underwater acoustic signal processing. It forms the foundation for the identification, localization, and tracking of underwater acoustic targets, aiming to acquire the directional information of targets from signals received by spatially distributed array elements. Unlike scalar hydrophones, which can only measure acoustic pressure, vector hydrophones are capable of simultaneously measuring acoustic pressure and particle velocity at the same point. Single vector hydrophones exhibit frequency-independent dipole directivity and a certain resistance to isotropic noise, enabling unambiguous full-space direction finding and effectively solving the problem of limited aperture in small underwater platform arrays. Therefore, in recent years, the research and application of single vector hydrophones in underwater acoustic DOA estimation have received widespread attention[1–4].

Common DOA estimation methods such as the Multiple Signal Classification (MUSIC)[5] algorithm and the Estimation of Signal Parameters via Rotational Invariance Techniques (ESPRIT)[6] algorithm have been proven to have high resolution and accuracy. These algorithms have laid the theoretical foundation for the application of DOA estimation using single vector hydrophones. Researchers have extensively explored this area: G.L. D'Spain analyzed and compared the beamforming results of single vector and vector arrays[7]; Wang et al. discussed the application of the Minimum Variance Distortionless Response (MVDR) beamforming technique to single differential vector hydrophone signal processing[8]; Tichavsky et al. proposed an ESPRIT algorithm based on a single vector hydrophone[9]; Levin proposed a DOA estimation method based on maximum likelihood estimation[10]; Liang et al. applied the MUSIC algorithm to single vector hydrophones and made improvements[11]; to solve the inconsistency problem of noise power between pressure and velocity channels, Liu et al. proposed a MUSIC algorithm that eliminates false sources[12]; Chen applied matrix filters to single vector hydrophones, enhancing the performance of the MUSIC algorithm[13].

Traditional methods rely on single vector hydrophones to measure the components of acoustic pressure and particle velocity for estimating the position of sound sources. However, these methods often face numerous challenges in complex environments, especially in multi-source environments and under noise interference. Due to the unique array orientation vector structure of single vector hydrophones and the fact that, even without channel amplitude errors, the noise power received by the pressure and velocity channels in isotropic noise environments is still different, the existing algorithms widely used for array DOA estimation are difficult to apply directly to single vector hydrophones[10–12].

Currently, sparse algorithms have demonstrated their superiority in the DOA domain by leveraging the sparse characteristics of spatial signals[14–18], effectively enhancing the accuracy and resolution of estimates even when the sparse distribution of signals is unknown[19–21]. In recent years, researchers have begun to explore the application of these algorithms to single vector hydrophones. Wang et al. applied the Sparse Asymptotic Minimum Variance (SAMV) algorithm to single vector hydrophones, achieving accurate estimation of target directions[22]. In recent years, gridless sparse methods such as atomic norm minimization[19,23–25] and gridless sparse iterative covariance-based estimation [16] method have proven their significant application potential in the field of array signal processing. The application of these methods has mainly been limited to Uniform Linear Arrays (ULAs) and Sparse Linear Arrays (SLAs), a limitation stemming from the specific mathematical structures they rely on, such as the Vandermonde decomposition of Toeplitz covariance matrices, which are readily satisfied in ULAs and SLAs[23–27]. The signal processing model of single vector hydrophones does not conform to this structure, leading to limitations in the application of such algorithms for DOA estimation with single vector hydrophones.

This paper aims to facilitate the application of such algorithms in single vector hydrophones by drawing on the practice of combined pressure and velocity channel processing in vector hydrophones. It investigates a Vector Signal Reconstruction (VSR) method for single vector hydrophones, which, through complex calculations and vector signal reconstruction, transforms the covariance matrix of single vector hydrophone signals into a Toeplitz structure suitable for gridless sparse methods. Furthermore, two sparse DOA estimation algorithms are proposed: the VSR Atomic Norm Minimization Based on Singular Value Decomposition (VSRANMSVD) algorithm and the VSR Structured Covariance Estimation (VSRSCE) algorithm. Compared to traditional methods, the proposed algorithms effectively improve the estimation accuracy and resolution in multi-source signals and low signal-to-noise ratio environments.

The main focus of this paper is to introduce the theoretical foundation and implementation details of the vector signal reconstruction method, and to verify its effectiveness in practical applications through simulation results. It further compares the significant advantages of the proposed algorithms over traditional DOA estimation methods in terms of directional estimation accuracy and resolution, especially highlighting their application potential in multi-source and low signal-to-noise ratio environments.

## II.  SINGLE VECTOR HYDROPHONE SIGNAL MODEL

Consider a mathematical model of a two-dimensional single vector hydrophone, which includes one acoustic pressure sensor and two vector velocity sensors that are perpendicular to each other on the horizontal plane, located along the $x$ and $y$ axes, respectively. Assume there are $K\,(K<3)$ far-field spatially uncorrelated narrowband signals $s_k(t)\,(k=1,...,K)$ impinging on the single vector hydrophone. The direction vector of the $k$ th signal $\boldsymbol{a}(\theta_k)=[1,\cos\theta_k,\sin\theta_k]^{\mathrm{T}}$ is a $3\times 1$ dimensional vector. Then, the received signal model can be represented as follows:

$$\boldsymbol{x}(t) = \begin{bmatrix} p(t) \\ v_x(t) \\ v_y(t) \end{bmatrix}$$
$$= \sum_{k=1}^{K} \boldsymbol{a}(\theta_k) s_k(t) + \boldsymbol{n}(t) \quad (1)$$
$$= \boldsymbol{A}(\vartheta) \boldsymbol{s}(t) + \boldsymbol{n}(t)$$

$\boldsymbol{x}(t)$ represents the signal received by the hydrophone at time $t$. Define the direction vector of the $k$ th signal $\boldsymbol{a}(\theta_k) = [1, \cos\theta_k, \sin\theta_k]^{\mathrm{T}}$, which is a three-dimensional column vector. The matrix $\boldsymbol{A}(\boldsymbol{\theta}) = [\boldsymbol{a}(\theta_1), ..., \boldsymbol{a}(\theta_K)]$ represents the direction vectors of all signals. The signal vector $\boldsymbol{s}(t) = [s_1(t), ..., s_K(t)]^{\mathrm{T}}$, and $\boldsymbol{n}(t)$ is a three-dimensional noise vector, which is assumed to be additive Gaussian white noise with zero mean. Furthermore, the signals and noise are assumed to be independent of each other. For DOA estimation, multiple snapshots (assuming $T$ snapshots) are typically used, and thus the received signal model can be described as follows:

$$\boldsymbol{X} = \boldsymbol{A}(\boldsymbol{\theta})\boldsymbol{S} + \boldsymbol{N} \quad (2)$$

$\boldsymbol{X} = [\boldsymbol{x}(1), \boldsymbol{x}(2), ..., \boldsymbol{x}(T)]$, $\boldsymbol{S} = [\boldsymbol{s}(1), \boldsymbol{s}(2), ..., \boldsymbol{s}(T)]$ and $\boldsymbol{N} = [\boldsymbol{n}(1), \boldsymbol{n}(2), ..., \boldsymbol{n}(T)]$.

The covariance matrix of $\boldsymbol{x}(t)$ can be represented as follows:

$$\boldsymbol{R}_x = \mathrm{E}[\boldsymbol{x}(t)\boldsymbol{x}^H(t)]$$
$$= \sum_{k=1}^{K} \sigma_{sk}^2 \boldsymbol{a}(\theta_k) \boldsymbol{a}(\theta_k)^H + E[\boldsymbol{n}(t)\boldsymbol{n}(t)^H]$$
$$= \boldsymbol{A}(\vartheta) \begin{bmatrix} \sigma_{s1}^2 & 0 & \cdots & 0 \\ 0 & \sigma_{s2}^2 & \cdots & \vdots \\ \vdots & \vdots & \ddots & 0 \\ 0 & \cdots & 0 & \sigma_{sk}^2 \end{bmatrix} \boldsymbol{A}^H(\vartheta) + \begin{bmatrix} \sigma_{np}^2 & 0 & 0 \\ 0 & \sigma_{nx}^2 & 0 \\ 0 & 0 & \sigma_{ny}^2 \end{bmatrix} \quad (3)$$
$$= \sum_{k=1}^{K} \sigma_{s_k}^2 \begin{bmatrix} 1 & \cos\theta_k & \sin\theta_k \\ \cos\theta_k & \cos^2\theta_k & \cos\theta_k \sin\theta_k \\ \sin\theta_k & \sin\theta_k \cos\theta_k & \sin^2\theta_k \end{bmatrix} + \begin{bmatrix} 1 & 0 & 0 \\ 0 & 1/2 & 0 \\ 0 & 0 & 1/2 \end{bmatrix} \sigma_n^2$$
$$= \boldsymbol{R}_s + \boldsymbol{R}_n$$

$\sigma_{sk}^2, k=1,2,...,K$ refers to the power of the $k$ th signal source. Meanwhile, $\sigma_{np}^2$, $\sigma_{nx}^2$, and $\sigma_{ny}^2$ represent the power of the noise in the acoustic pressure channel, velocity $x$ channel, and velocity $y$ channel, respectively. The covariance matrix of the signals $\boldsymbol{R}_s$ and the covariance matrix of the noise $\boldsymbol{R}_n$ can be represented accordingly as:

$$\boldsymbol{R}_s = \sum_{k=1}^{K} \sigma_{s_k}^2 \begin{bmatrix} 1 & \cos\theta_k & \sin\theta_k \\ \cos\theta_k & \cos^2\theta_k & \cos\theta_k\sin\theta_k \\ \sin\theta_k & \sin\theta_k\cos\theta_k & \sin^2\theta_k \end{bmatrix}$$

$$\boldsymbol{R}_n = \begin{bmatrix} 1 & 0 & 0 \\ 0 & 1/2 & 0 \\ 0 & 0 & 1/2 \end{bmatrix} \sigma_n^2 \tag{4}$$

## III. SPARSE DOA METHOD BASED ON VSR

### A. VSR Method

Upon reexamining the structure of the signal covariance matrix $\boldsymbol{R}_s$, it is observed that this matrix does not fully meet all the requirements of a Toeplitz structure. A key characteristic of Toeplitz matrices is that the elements on each diagonal should be identical, which is not entirely the case with $\boldsymbol{R}_s$. The definition of $\boldsymbol{R}_s$ is as follows:

$$\boldsymbol{R}_s = \sum_{k=1}^{K} \sigma_{s_k}^2 \begin{bmatrix} 1 & \cos\theta_k & \sin\theta_k \\ \cos\theta_k & \cos^2\theta_k & \cos\theta_k\sin\theta_k \\ \sin\theta_k & \sin\theta_k\cos\theta_k & \sin^2\theta_k \end{bmatrix} \tag{5}$$

In $\boldsymbol{R}_s$, although the off-diagonal elements are equal to the elements at symmetrical positions along the diagonals, showing a certain level of symmetry, the diagonal elements (such as $1$, $\cos^2\theta_k$, and $\sin^2\theta_k$) are not consistent and do not fully comply with the definition of a Toeplitz matrix. In this study, a vector signal reconstruction algorithm is proposed, which processes the covariance matrix of signals received by a single vector hydrophone to satisfy the Toeplitz structure through a specific transformation process. The transformation is given by the following equation, where $\boldsymbol{y}(t)$ represents the reconstructed signal:

$$\boldsymbol{y}(t) = \begin{bmatrix} v_x(t) - j \cdot v_y(t) \\ p(t) \\ v_x(t) + j \cdot v_y(t) \end{bmatrix}$$

$$= \begin{bmatrix} \sum_{k=1}^{K} \cos(\theta_k) s_k(t) - j \sum_{k=1}^{K} \sin(\theta_k) s_k(t) \\ \sum_{k=1}^{K} s_k(t) \\ \sum_{k=1}^{K} \cos(\theta_k) s_k(t) + j \sum_{k=1}^{K} \sin(\theta_k) s_k(t) \end{bmatrix} + \begin{bmatrix} n_x(t) - j \cdot n_y(t) \\ p(t) \\ n_x(t) + j \cdot n_y(t) \end{bmatrix} \qquad (6)$$

Equation(6) can be obtained by left multiplying $\boldsymbol{x}(t)$ by the vector signal reconstruction matrix $\boldsymbol{G}$, which can be specifically represented as:

$$\boldsymbol{G} = \begin{bmatrix} 0 & 1 & -j \\ 1 & 0 & 0 \\ 0 & 1 & j \end{bmatrix} \qquad (7)$$

By applying Euler's formula $e^{j\theta} = \cos(\theta) + j\sin(\theta)$ and $e^{-j\theta} = \cos(\theta) - j\sin(\theta)$ to Equation (6), the equation can be further simplified to obtain the following expression:

$$\boldsymbol{y}(t) = \sum_{k=1}^{K} \begin{bmatrix} e^{-j\theta_k} s_k(t) \\ s_k(t) \\ e^{j\theta_k} s_k(t) \end{bmatrix} + \begin{bmatrix} n_x(t) - j \cdot n_y(t) \\ p(t) \\ n_x(t) + j \cdot n_y(t) \end{bmatrix} \qquad (8)$$

Through vector signal reconstruction, the covariance matrix of data received by a single vector hydrophone can be represented as:

$$\begin{aligned}
\boldsymbol{R}_y &= \mathrm{E}[\boldsymbol{y}(t)\boldsymbol{y}^H(t)] \\
&= \sum_{k=1}^{K} \begin{bmatrix} e^{-j\theta_k} s_k(t) \\ s_k(t) \\ e^{j\theta_k} s_k(t) \end{bmatrix} \begin{bmatrix} e^{-j\theta_k} s_k(t) \\ s_k(t) \\ e^{j\theta_k} s_k(t) \end{bmatrix}^H + \begin{bmatrix} n_x(t) - j \cdot n_y(t) \\ p(t) \\ n_x(t) + j \cdot n_y(t) \end{bmatrix} \begin{bmatrix} n_x(t) - j \cdot n_y(t) \\ p(t) \\ n_x(t) + j \cdot n_y(t) \end{bmatrix}^H \\
&= \sum_{k=1}^{K} \sigma_{s_k}^2 \begin{bmatrix} 1 & e^{-j\theta} & e^{-j2\theta} \\ e^{j\theta} & 1 & e^{-j\theta} \\ e^{j2\theta} & e^{j\theta} & 1 \end{bmatrix} + \sigma_n^2 \begin{bmatrix} 1 & 0 & 0 \\ 0 & 1 & 0 \\ 0 & 0 & 1 \end{bmatrix}
\end{aligned} \quad (9)$$

Furthermore, $\boldsymbol{R}_y$ can be simplified as:

$$\boldsymbol{R}_y = \boldsymbol{\Phi}(\boldsymbol{\theta}) \mathrm{diag}(\boldsymbol{\sigma}_s^2) \boldsymbol{\Phi}^H(\boldsymbol{\theta}) + \mathrm{diag}(\boldsymbol{\sigma}_n^2) \quad (10)$$

Where, $\boldsymbol{\sigma}_s^2 = [\sigma_{s1}^2, \sigma_{s2}^2, ..., \sigma_{sK}^2]^H$ and $\boldsymbol{\sigma}_n^2 = [\sigma_n^2, \sigma_n^2, ..., \sigma_n^2]^H$ represent the power vectors of signals and noise respectively. $\boldsymbol{\Phi}(\boldsymbol{\theta})$ is the steering vector of the single vector signal after vector signal reconstruction, which can be represented by $\boldsymbol{G}$ as:

$$\begin{aligned}
\boldsymbol{\Phi}(\boldsymbol{\theta}) &= \boldsymbol{G}\boldsymbol{A}(\boldsymbol{\theta}) \\
&= \begin{bmatrix} 0 & 1 & -j \\ 1 & 0 & 0 \\ 0 & 1 & j \end{bmatrix} \begin{bmatrix} 1 & 1 & ... & 1 \\ \cos(\theta_1) & \cos(\theta_2) & ... & \cos(\theta_K) \\ \sin(\theta_1) & \sin(\theta_2) & ... & \sin(\theta_K) \end{bmatrix} \\
&= \begin{bmatrix} e^{-j\theta_1} & e^{-j\theta_2} & ... & e^{-j\theta_K} \\ 1 & 1 & ... & 1 \\ e^{j\theta_1} & e^{j\theta_2} & ... & e^{j\theta_K} \end{bmatrix} \\
&= [\boldsymbol{\Phi}(\theta_1), \boldsymbol{\Phi}(\theta_2), ..., \boldsymbol{\Phi}(\theta_K)]
\end{aligned} \quad (11)$$

In this processing step, by performing vector signal reconstruction on the single vector received data, the steering vector is approximated as a three-element uniform linear array. This method not only retains the core characteristics of the original signal but also provides an appropriate mathematical framework for subsequent signal processing and analysis tasks. When using $T$ snapshots for Direction of Arrival (DOA) estimation, the vector signal reconstruction model of the received signal $\boldsymbol{Y} = [\boldsymbol{y}(1), \boldsymbol{y}(2), ..., \boldsymbol{y}(T)]$ can be represented as follows:

$$\boldsymbol{Y} = \boldsymbol{\Phi}(\boldsymbol{\theta})\boldsymbol{S} + \boldsymbol{G}\boldsymbol{N} \quad (12)$$

## B. Atomic Norm Minimization Based on Singular Value Decomposition

The previous section introduced the basic principles and implementation methods of VSR. This method ensures that the covariance matrix of the single vector received signals possesses a Toeplitz structure, which is a prerequisite for employing atomic norm minimization. Atomic norm minimization(ANM) aims to achieve DOA estimation by minimizing the number of atoms that constitute the signal. The regularized optimization problem can be formulated as:

$$\min_{Z} \lambda \mathcal{M}(Z) + \frac{1}{2} \|Z - Y\|_F^2, \tag{13}$$

Where, $\lambda > 0$ represents the regularization parameter, and $\mathcal{M}(Z)$ denotes the sparsity measure. By minimizing $\mathcal{M}(Z)$, we aim to reduce the number of atoms constituting $Z$. In the presence of noise, the ANM problem further transforms into the following Semidefinite Programming (SDP) problem:

$$\min_{Q,u,Z} \frac{\lambda}{2\sqrt{N}} [\mathrm{Tr}(Q) + \mathrm{Tr}(T(u))] + \frac{1}{2} \|Z - Y\|_F^2,$$
$$\text{subject to} \begin{bmatrix} Q & Z^H \\ Z & T(u) \end{bmatrix} \geq 0 \tag{14}$$

To select an appropriate $\lambda$ parameter, set $\lambda \approx 3(T + \log 3 + \sqrt{2T\log 3})\sigma_n$. When there are many snapshots, the computational load of ANM is significant. In such cases, Singular Value Decomposition (SVD) is used to simplify the data:

$$Y = ULV^H \tag{15}$$

A $3 \times L$ dimensional data matrix $Y_{\mathrm{SV}} = ULD_L^T = YVD_L^T$ is defined, where $Y_{\mathrm{SV}}$ contains most of the signal energy. Here, $D_L = [I_L, 0]$, and $I_L$ is a $K$-order identity matrix. Therefore, the optimization problem can be reformulated as:

$$\min_{Q_{\text{SV}},u,Z_{\text{SV}}} \frac{\lambda}{2\sqrt{N}} \left[ \text{Tr}(Q_{\text{SV}}) + \text{Tr}(T(u)) \right] + \frac{1}{2} \| Q_{\text{SV}} - Y_{\text{SV}} \|_F^2,$$
$$\text{subject to} \begin{bmatrix} Q_{\text{SV}} & Z_{\text{SV}}^H \\ Z_{\text{SV}} & T(u) \end{bmatrix} \geq 0 \tag{16}$$

By solving Equation (16), the optimized covariance matrix $R_{\text{SV}} = Z_{\text{SV}} Z_{\text{SV}}^H$ is obtained, which further allows the estimation of the target direction to be addressed through the MUSIC algorithm.

### C. Structured Covariance Estimation

Under noise-free conditions, the covariance matrix of the $\boldsymbol{y}(t)$ signal in Equation (9) can be represented as:

$$\Sigma^\star = \boldsymbol{\Phi}(\boldsymbol{\theta}) \text{diag}(\boldsymbol{\sigma}_s^2) \boldsymbol{\Phi}^H(\boldsymbol{\theta}) \tag{17}$$

$\Sigma^\star$ is a positive semidefinite (PSD) Hermitian Toeplitz matrix. This matrix exhibits low-rank characteristics, with its rank $\text{rank}(\Sigma^\star) = r < 3$. At this point, the sparsity of the target bearings translates into the low-rank property of the covariance matrix. When the number of snapshots is $T$, the covariance matrix of the single vector received data can be given by the following formula:

$$\begin{aligned} \Sigma &= \mathbb{E}[YY^H] \\ &= \frac{1}{T} YY^H \\ &= \frac{1}{T} \sum_{t=1}^{T} y(t) y^*(t) \end{aligned} \tag{18}$$

The next step involves finding a low-rank positive semidefinite (PSD) Hermitian Toeplitz matrix. The optimization problem can be represented as:

$$\min_{u} \frac{1}{2} \| T(u) - \Sigma \|_F^2 + \tau \text{rank}(T(u)), \text{subject to } T(u) \succeq 0, \tag{19}$$

Where $\tau$ is a regularization parameter used to balance the data fitting term and the rank regularization term. Since directly minimizing the rank is an NP-hard problem, a convex relaxation on the positive semidefinite (PSD) cone of the rank minimization problem is required. This involves replacing rank minimization with trace minimization, resulting in:

$$\min_{\boldsymbol{u}} \frac{1}{2} \| \mathcal{T}(\boldsymbol{u}) - \boldsymbol{\Sigma} \|_F^2 + \tau \operatorname{Tr}(\mathcal{T}(\boldsymbol{u})), \quad \text{subject to } \mathcal{T}(\boldsymbol{u}) \succeq 0. \tag{20}$$

$\tau = 2.5 \times 10^{-3} / ((\log T)^2 \log 3)$, the SCE algorithm utilizes the sample covariance matrix $\boldsymbol{\Sigma}$ instead of directly using $\boldsymbol{Y}$. In the case of a two-dimensional single vector hydrophone, it does not require storing the $3 \times T$ matrix $\boldsymbol{Y}$, but only the $3 \times 3$ matrix $\boldsymbol{\Sigma}$, reducing computational load when there are a large number of snapshots. By solving the semidefinite programming problem, the optimized matrix $\boldsymbol{R}_{SCE}$ of the structured covariance matrix algorithm is obtained, with $\boldsymbol{R}_{SCE} = \mathcal{T}(\boldsymbol{u}^*)$, where $\boldsymbol{u}$ is solved by (20). Further, the target bearings are solved using the MUSIC algorithm.

## IV. SIMULATION ANALYSIS

In the scenario where a single vector hydrophone receives signals from $K$ far-field targets, the signal-to-noise ratio (SNR) for the $k$th source in a Gaussian noise environment is defined as follows:

$$\operatorname{SNR} = 10 \log \left( \frac{E[|x_k(t)|^2]}{\sigma} \right), \quad k = 1, ..., K. \tag{21}$$

The Root Mean Square Error (RMSE) regarding the accuracy of DOA estimation is defined as:

$$\operatorname{RMSE} = \sqrt{\frac{1}{KZ} \sum_{k=1}^{K} \sum_{z=1}^{Z} \left( \hat{\theta}_k^z - \theta_k^z \right)^2}, \tag{22}$$

Here, $\theta_k^z$ represents the true DOA of the $k$th source in the $z$th Monte Carlo simulation, while $\hat{\theta}_n^z$ is the estimated value of $\theta_k^z$. Assuming the number of sources $K$ is known, the positions of the $K$ largest peaks are selected as the estimated DOA values $\{\hat{\theta}_k^z\}_{k=1}^K$, and the RMSE is calculated based on the average results from $Z = 400$ runs.

### A. Single-Target Simulation

In this section, a single-target scenario is considered, where the incident angle of the target is $-30°$, with the background noise being complex Gaussian white noise unrelated to the signal. In the simulation, the number of snapshots is set to 1000, with SNR designated at 10dB and 0dB respectively. The azimuth search step is set to $1°$. This simulation compares the performance of Conventional Beamforming (CBF), Minimum Variance Distortionless Response (MVDR), MUSIC, the Iterative Adaptive Approach (IAA) [28], SParse Iterative Covariance-based Estimation (SPICE) [29], SPICE+ [29] algorithms, VSRANMSVD algorithm, and the VSRSCE algorithm, as illustrated in Figure 1. The vertical red dashed line represents the actual direction of incidence of the target, and all algorithms effectively estimated the target's bearing. Particularly, the CBF algorithm shows the widest main lobe width, while the VSRANMSVD and VSRSCE algorithms demonstrate narrower main lobe widths relative to other algorithms, with lower sidelobe levels. The VSRSCE algorithm exhibits the lowest sidelobe level, showing the sharpest peak, which is more pronounced at an SNR of 0dB.

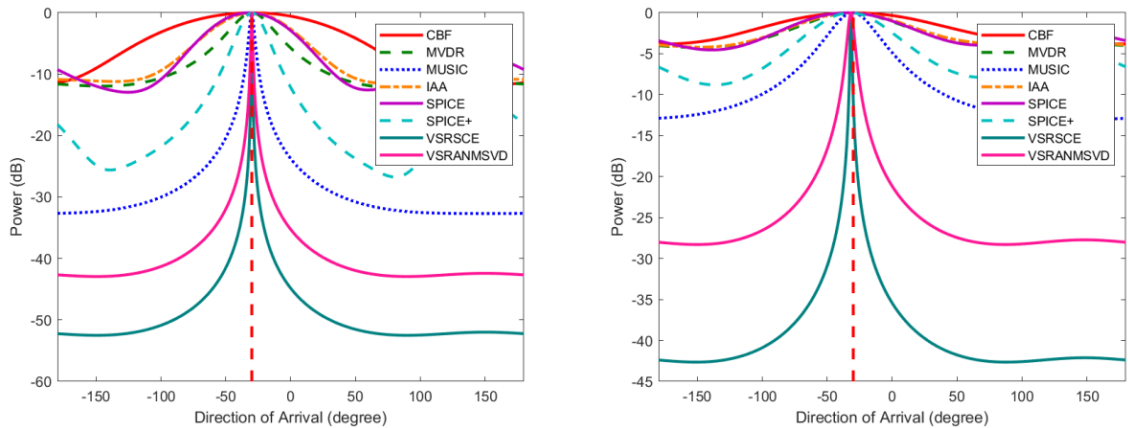

(a)                                      (b)

FIG. 1.Spatial Spectrum Graphs of Different DOA Algorithms for a Single Target:(a) SNR=10dB (b) SNR=0dB)

Figure 2 displays the RMSE curves of different DOA estimation algorithms at various SNRs. In the context of a single target, the RMSE of azimuth estimation for all the mentioned algorithms decreases with an increase in the SNR. Below an SNR of -5dB, the RMSE rapidly increases with the improvement of signal SNR. When the SNR is greater than -5dB, except for the SPICE algorithm, the RMSE of azimuth estimation for the mentioned algorithms is less than $2°$. When the SNR is above 0dB, excluding the SPICE algorithm, the RMSE of azimuth estimation for other methods is essentially consistent. Below an SNR of 0dB, the RMSEs of CBF, MVDR, IAA, SPICE+, VSRANMSVD, and VSRSCE algorithms are basically the same, with the MUSIC algorithm's RMSE slightly higher than these algorithms.

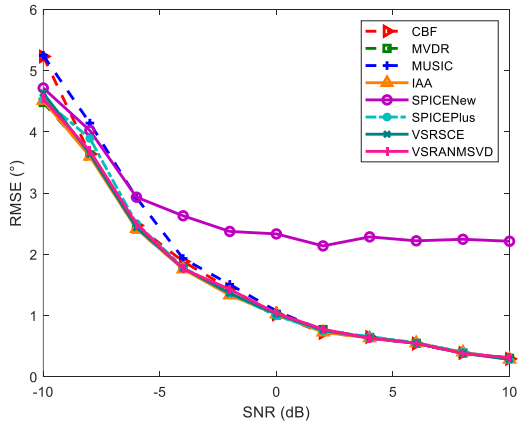

FIG. 2.Comparative Analysis of RMSE Performance for a Single-Target Scenario

**B. Dual-Target Simulation**

Next, consider a scenario with two targets, where the incident angles of the targets are $\theta_1 = -30°$ and $\theta_2 = 20°$, with the background noise being complex Gaussian white noise unrelated to the signal. In the simulation, the number of snapshots is set to 1000, with Signal-to-Noise Ratios (SNRs) of 10dB and 0dB. The azimuth search step is $1°$. As shown in Figure 3, in the dual-target scenario, the azimuth spectra of CBF, MVDR, IAA, SPICE, and SPICE+ algorithms only display a single peak, indicating that under these simulation conditions, these algorithms cannot differentiate between the two targets. However, under the condition of an SNR of 10dB, the MUSIC, VSRANMSVD, and VSRSCE algorithms are all capable of accurately estimating the bearings of the targets, and the VSRSCE algorithm exhibits sharper peaks. When the SNR drops to 0dB, the estimation results of the MUSIC algorithm show a significant deviation, which significantly differ from the preset target bearings. Meanwhile, the VSRANMSVD and VSRSCE algorithms are still able to accurately estimate the bearings of the targets.

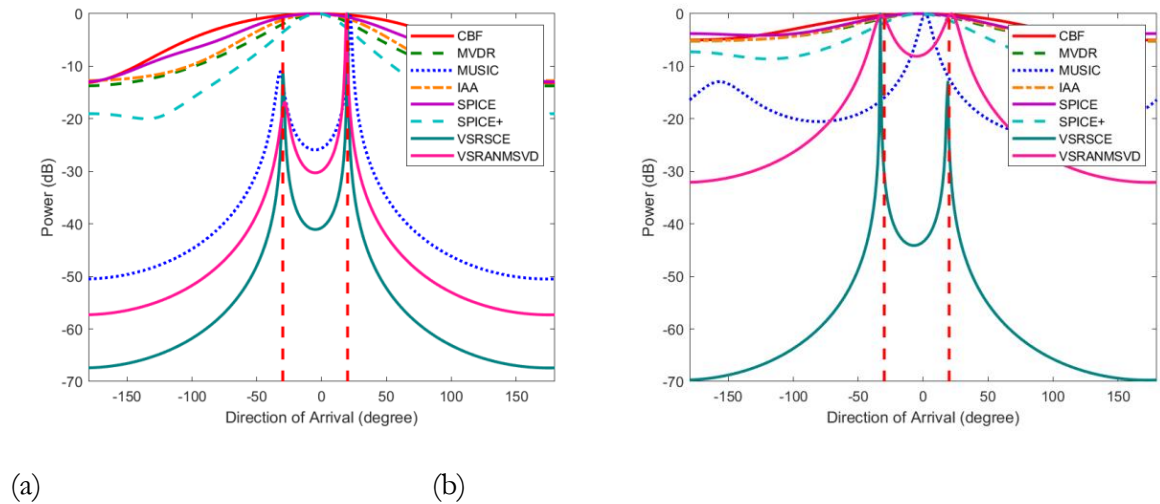

(a)　　　　　　　　　　　　(b)

FIG. 3. Spatial Spectrum Estimation Comparison of Various Algorithms: (a) SNR=10dB (b) SNR=0dB

From the spatial spectra mentioned above, it is observed that the CBF, MVDR, MUSIC, IAA, SPICE, and SPICE+ algorithms cannot differentiate between the two targets. Therefore, the following images will not compare with the aforementioned algorithms. Figure 4 illustrates the relationship between the Root Mean Square Error (RMSE) of azimuth estimation and Signal-to-Noise Ratio (SNR) for the MUSIC, VSRANMSVD, and VSRSCE algorithms in a dual-target scenario. The MUSIC algorithm is represented by a blue dashed line. Below an SNR of -2dB, the MUSIC algorithm cannot differentiate between the two targets, hence the curve is not plotted below SNR=-2dB. As the SNR increases, its RMSE significantly decreases, indicating an improvement in estimation accuracy. Especially when the SNR increases from -10 dB to about 2 dB, the RMSE of the MUSIC algorithm drops rapidly, then the decline trend slows down but continues to decrease.The VSRANMSVD and VSRSCE algorithms overall exhibit lower RMSE across all SNR levels compared to the MUSIC algorithm, indicating better performance within this test range. With the increase of SNR, the RMSE of both VSRANMSVD and VSRSCE algorithms also shows a declining trend, and their RMSE drops to near 0° at high SNR values, showing higher localization accuracy. The VSRSCE algorithm performs better across the entire SNR range relative to the VSRANMSVD and MUSIC algorithms.

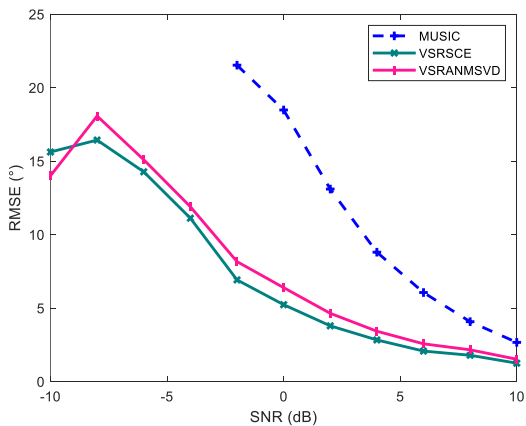

FIG. 4.Comparative Analysis of RMSE Performance for a Dual-Target Scenario

Figure 5 shows the performance of MUSIC, VSRANMSVD, and VSRSCE algorithms in terms of resolution probability at different Signal-to-Noise Ratio (SNR) levels. The horizontal axis represents the SNR, measured in decibels (dB), ranging from -10 dB to 10 dB. The vertical axis represents the resolution probability, from 0 to 1. In the figure, the MUSIC algorithm is represented by a blue dashed line, and its performance improves with an increase in SNR, but the performance growth is slower under low SNR conditions. Compared to the MUSIC algorithm, the VSRANMSVD and VSRSCE algorithms exhibit higher resolution probabilities across the entire SNR range. Furthermore, the VSRSCE algorithm shows superior performance, especially when SNR is greater than -4dB, approaching 1, indicating better performance than the MUSIC algorithm.

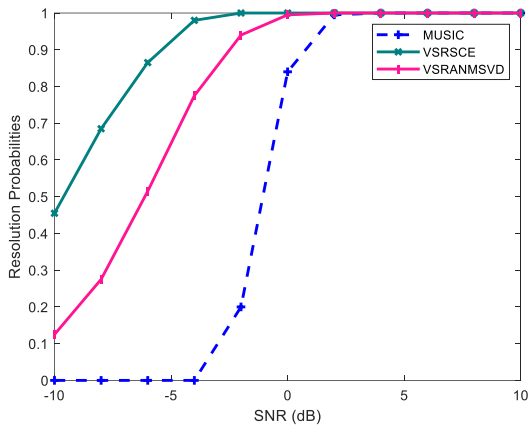

FIG. 5.Comparative Analysis of Resolution Probability for a Dual-Target Scenario

## V. CONCLUSION

In this study, we explored the application of single vector hydrophones in underwater acoustic DOA estimation. Addressing the challenges posed by multi-source signals and noise interference faced by traditional DOA estimation methods, this research proposed two sparse DOA algorithms based on single vector hydrophones, namely VSRANMSVD and VSRSCE algorithms. Through theoretical analysis and simulation experiments, we demonstrated that these two algorithms could effectively enhance the accuracy and resolution of DOA estimation, particularly outperforming traditional methods in low SNR and multi-source environments.